\documentstyle[twocolumn,aps,prl,epsfig,psfig]{revtex}
\begin{document}


\title{
Nuclear Structure Functions in the
  Large ${\bf{x}}$  Large ${\bf{Q^2}}$ Kinematic Region in Neutrino
  Deep Inelastic Scattering}

\author{ M. ~Vakili$^3$, 
  	C.~G.~Arroyo$^4$, P.~Auchincloss$^9$, L.~de~Barbaro$^7$, 
	P.~de~Barbaro$^9$, 
	A.~O.~Bazarko$^4$, R.~H.~Bernstein$^5$, A.~Bodek$^9$, 
	T.~Bolton$^6$, H.~Budd$^9$, J.~Conrad$^4$, 
	D.~A.~Harris$^9$, R.~A.~Johnson$^3$,
	J. ~H.~Kim$^4$, B.~J.~King$^4$,
	T.~Kinnel$^{10}$, G.~Koizumi$^5$, S.~Koutsoliotas$^4$, 
	M.~J.~Lamm$^5$, W.~C.~Lefmann$^1$, W.~Marsh$^5$, 
	K.~S.~McFarland$^9$, C.~McNulty$^4$, S.~R.~Mishra$^4$, 
	D.~Naples$^5$, P.~Nienaber$^{11}$,
	M.~J.~Oreglia$^2$, L.~Perera$^3$, P.~Z.~Quintas$^4$, 
	A.~Romosan$^4$, W.~K.~Sakumoto$^9$, B.~A.~Schumm$^2$, 
	F.~J.~Sciulli$^4$, W.~G.~Seligman$^4$, M.~H.~Shaevitz$^4$, 
	W.~H.~Smith$^{10}$, P.~Spentzouris$^4$, R.~Steiner$^1$, 
	E.~G.~Stern$^4$, U.~K.~Yang$^9$, J.~Yu$^5$ }

\address{(1)~Adelphi University, Garden City, NY 11530 USA; \\
(2)~University of Chicago, Chicago, IL 60637 USA; \\
(3)~University of Cincinnati, Cincinnati, OH 45221 USA;\\ 
(4)~Columbia University, New York, NY 10027 USA; \\
(5)~Fermi National Accelerator Laboratory, Batavia, IL 60510 USA; \\
(6)~Kansas State University, Manhattan, KS 66506 USA;\\
(7)~Northwestern University, Evanston, IL 60208 USA;\\
(8)~University of Oregon, Eugene, OR 97403 USA;\\
(9)~University of Rochester, Rochester, NY 14627 USA;\\
(10)~University of Wisconsin, Madison, WI 53706 USA;\\
(11)~Xavier University, Cincinnati, OH 45207 USA }

\date{\today}



\address{~
\parbox{15cm}{\rm
\medskip
Data from the CCFR E770  Neutrino Deep Inelastic Scattering (DIS)
 experiment at Fermilab contain events with large Bjorken $x$ ($x>0.7$) and
 high momentum transfer ($Q^2>50\ ({\rm GeV/c})^2$).
A comparison of the data with a model based on no
nuclear effects at large $x$,\ shows a significant 
excess of events in the data.
Addition of Fermi gas motion of the nucleons in the nucleus to the
 model does not explain the excess. Adding a higher momentum
 tail due to the formation of ``quasi-deuterons'' 
makes some improvement. 
An exponentially falling $F_2 \propto e^{-s(x-x_0)}$ at large $x$,
predicted by ``multi-quark clusters'' and ``few-nucleon
correlations'', can describe the data. A value of  
 $s=8.3 \pm 0.7(stat.)\pm 0.7(sys.)$ yields the best agreement with the
 data. 
}}

\maketitle

Deep inelastic scattering (DIS) neutrino interactions are ideally suited for
measuring the nuclear structure functions over a wide range of kinematic 
variables.  
The CCFR experiment has collected over $10^6$ charged current events 
where neutrinos and antineutrinos scattered off of
an iron target.  
We have recently published\cite{bill} structure functions
for the Bjorken $x$ region of $x \le 0.75$ based on these data.  In this
paper, we extend the measurement into the $x > 0.75$ region\cite{masoud}.
We have approximately 2000 events with $x>0.75$.

In the infinite momentum frame, $x$ is 
the fraction of the nucleon's momentum carried by the struck quark.
As such, $x$ is kinematically constrained to be less than 1.  In the 
same model, quark counting rules\cite{counting}
would imply that the quark structure
functions should decrease like $(1-x)^3$ near $x=1$.

When $x$ is measured in DIS experiments, it is assumed that the incident
lepton scatters off a nucleon at rest in the laboratory.  Under this
assumption and neglecting terms proportional to the nucleon's mass, 
$x$ takes on the form
\begin{equation}
x= {2 E_{l} E_{l^\prime}\over  M_N E_{had}} \sin^2 {\theta\over 2} \ ,
\end{equation}
where $E_{l}$ is the energy of the incoming lepton, $E_{l^\prime}$ is the
energy of the outgoing lepton, $\theta$ is the laboratory angle between the
two leptons, $E_{had}$ is the energy of the outgoing 
hadron shower, and $M_N$ is the mass of the struck nucleon.
However, for nuclear targets (like the
iron detector of CCFR) the nucleons need not be at rest in the laboratory.
The nucleons themselves have Fermi motion.
Furthermore, the quarks inside a given nucleon may be
affected by the surrounding nucleons either through the exchange of mesons
inside the nucleus, through the formation of few-nucleon states, or
through the condensation of the nucleons into multi-quark clusters
(``bags'').  All of these phenomena have been used to explain the
EMC effect\cite{EMC}.  
Nuclear motion and quark
interactions ``smear'' out the nucleon's structure functions, 
taking events from one region in $x$ and $Q^2$ and moving
them into adjacent regions.
In the regions below $x=0.75$, the differences between the nuclear and 
nucleon structure functions are 
small mainly because the nucleon structure functions are slowly varying. 
At $x=0.7$, the nuclear correction is expected to be about 10\%\cite{bill}.
However, in the higher $x$ region, nucleon structure functions
must go to zero while the nuclear structure functions need not.
In fact, most of the cross section
near $x=1$ and all of the cross section for $x>1$ comes from the
fact that the quark exists in the nuclear environment.   
Thus, the high $x$ region is ideally suited for investigating
nuclear effects in DIS scattering.


Event selection in this
analysis is very similar to previous CCFR analyses\cite{bill,CCFRs}.  
However, because there are so few true high $x$ events, we are
very susceptible to background events from the lower $x$ region being
mismeasured and appearing as high $x$ events.
To minimize this effect, we further restrict the kinematic region for
this analysis over previous CCFR analyses 
to the region where $x$ is well measured.
The resolution of $x$ is poor at large muon energies,
at small hadron energies, and at small muon angles.
We also require that the hadronic energy is well
contained in the calorimeter and has a small measurement error and that 
the outgoing muon traverses the whole length of
the toroid spectrometer.  
For these reasons, we require events to be in the kinematic region  
 $15\ {\rm GeV} <E_\mu < 360\ {\rm GeV}$, $20<E_{had} <360\ {\rm GeV}$,
 $Q^2 < 400\ ({\rm GeV/c})^2$, and $\theta_\mu > 17\ {\rm mrad}$.
To minimize the uncertainties from higher twist contributions and from
resonant scattering, we accept events where 
$Q^2 > 50\ ({\rm GeV/c})^2  $.  The average $Q^2$ of the accepted events 
is 120 (GeV/c)$^2$.

In addition to these  quality cuts, we made various checks on the muon 
momentum and hadronic energy measurements. We compared different 
measurement algorithms and hand-scanned  all of the high $x$ events. 
There were about 30 events with an apparent discrepancy between two different
hadron energy or muon energy measurement algorithms, 
indicating possible mismeasurements. These events were removed
from the sample.

\begin{figure}[t]
\begin{center}
\mbox{\epsfxsize7.5cm\epsfxsize8.5cm\epsffile{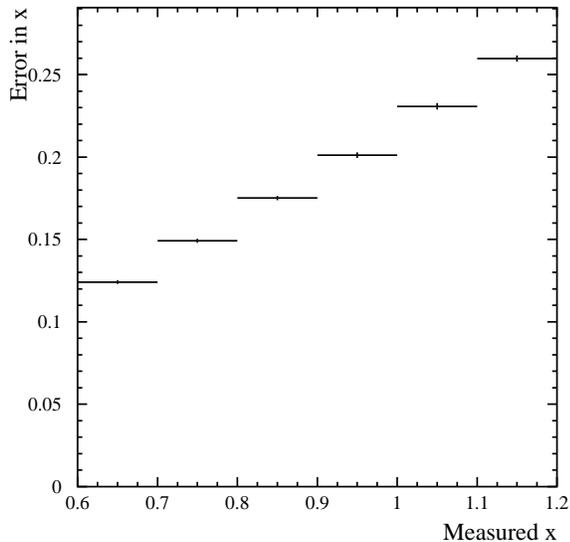}}
\end{center}
\caption{
Error in $x$ as a function of measured $x$ in the data.
}
\label{fig:resolution}
\end{figure}

As with any steeply falling function, the measured distribution
is distorted by
measurement resolution.
In CCFR, the hadron and muon energy resolution functions 
are measured
from test beam data and from internal calibrations\cite{res}.
They are well known over four
orders of magnitude.  
An extensive Monte Carlo program has been written to simulate the effects
of the detector on neutrino interactions.  With this program, we 
predict the observed $x$ distribution of a given input model. For
models with free parameters, we vary the free parameter until we get
the best agreement between the predicted $x$ distribution and that 
observed.  The $x$ resolution as a function of $x$ is 
shown in Fig.~\ref{fig:resolution}.  The number plotted is the resolution averaged over the 
accepted Monte Carlo events.

Neutrino DIS can be completely parameterized (up
to effects that are proportional to the muon mass over the neutrino
energy) by three structure functions: $2 x F_1$, $F_2$, and $x F_3$.  In the
quark parton model, $2 x F_1$ and $F_2$ differ by the distribution of 
effective scalar partons in the nucleon which arise from the 
apparent transverse
momentum of the quarks.  At high $Q^2$,
the scalar effects are expected to be insignificant.
$2 x F_1$ and $x F_3$ differ by a contribution 
from sea quarks.  At high $x$, the sea quark contribution to the structure 
functions is also negligible.  Therefore, for this analysis, we assume 
that\cite{masoud}, for $x>0.6$, 
\begin{equation}
2 x F_1(x,Q^2) = F_2(x,Q^2) = x F_3(x,Q^2).\label{sf}
\end{equation}

\setlength{\unitlength}{1.0mm}
\begin{figure}[t]
\begin{picture}(80,75)(0,1)
\mbox{\psfig{figure=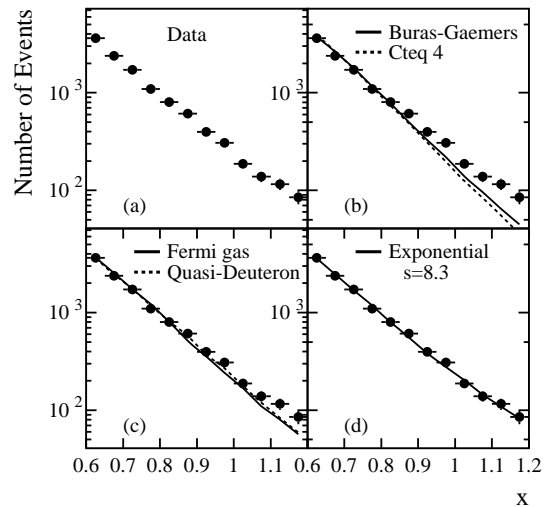,width=8.5cm,bbllx=.25in,bblly=2.25in,bburx=7.5in,bbury=9.25in}}
\end{picture}
\caption{ 
(a) Measured $x$ distribution. 
Comparison of measured $x$ distribution and 
(b) distribution 
predicted by Buras-Gaemers structure functions and by CTEQ4M structure 
functions,
(c) distributions predicted by a flat Fermi gas model and by the 
Bodek-Ritchie model,
(d) distributions with an exponentially falling $F_2$ with s=8.3.
All error bars represent only the  statistical errors.
}
\label{fig:models}
\end{figure}

As stated above, simple nucleon models of 
neutrino-iron
scattering do not include the nuclear environment and therefore
are not expected to describe our $x$ distributions.  However, it is instructive
to compare a model including only nucleon structure functions with our data
to see the difference. 
Fig.~\ref{fig:models}(a) shows the measured $x$ distribution from the data. In
Fig.~\ref{fig:models}(b), we show our expected 
$x$ distribution generated from the Buras-Gaemers (BG)\cite{BG} and
from the CTEQ4M parameterizations of the structure functions\cite{CTEQ}. 
 The former have been fit to our $x<0.7$
data and, as such, include the nuclear effects that are seen at lower $x$.
Likewise, the latter uses CCFR data in their fits and therefore, to some
extent, must include nuclear effects.
However, both parameterizations are approximately 
proportional to $(1-x)^3$ as $x$ goes to 1
and therefore do not include high $x$ nuclear effects.  
As can be seen in Fig.~\ref{fig:models}(b), the BG and CTEQ4M parameterizations  
seriously underestimate the high $x$ cross section.


When a nuclear Fermi gas model with a flat momentum distribution up to a 
Fermi surface of 257 MeV/c is added to the
 BG parameterization, the result again underestimates
the high $x$ measurement 
as shown in Fig.~\ref{fig:models}(c). It has been found in
low energy electron-nucleus scattering measurements and in analyzing the EMC
effect in high energy experiments that a high momentum tail has to be
added to the pure Fermi gas distribution to reproduce the data. 
The model of Bodek and Richie\cite{Bodek} introduces the high momentum
tail through the inclusion of quasi-deuteron scattering.
Fig.~\ref{fig:models}(c)  compares this model with our data.  
While the 
comparison with our data is still not perfect, it is slightly better than 
that for the models with no nucleon motion or with only Fermi motion.


After the discovery of the EMC effect, various more exotic nuclear
 models were proposed.
One such model is  the 
few-nucleon correlation model of
Frankfurt and Strikman\cite{strickman}. 
Other authors such as Kondratyuk and Shmatikov\cite{bags} suggest 
models where the neutrinos scattered off 
higher quark-count states (6-quark, 9-quark, etc.).  In such models
the individual quarks can have a higher effective momentum than in the
isolated nucleon.  Both the few-nucleon correlation models and the 
multi-quark cluster models can be parameterized as an exponentially falling 
$F_2$ structure function ($F_2(x) \propto e^{-s x}$) in the large $x$ region. 
We have used the parameterized structure
function from the BCDMS collaboration\cite{BCDMS} and varied the 
exponential slope parameter $s$ to minimize the $\chi^2$
difference between the Monte Carlo prediction and the measured distributions
in the region $0.6<x<1.2$.  
We find that $s =  8.3$ minimizes the 
$\chi^2$ at 8.2 for 12 degrees of freedom.  
The region that changes the total $\chi^2$ by 1 unit leads
to an error estimate of $\delta s= \pm 0.7$.  
Fig.~\ref{fig:models}(d) compares our best exponential fit
to the data.  Fig.~\ref{fig:f2} shows $F_2$ with its error band
from this exponential model normalized to the $x=0.65$ point from the
CCFR structure functions\cite{bill}.

\setlength{\unitlength}{1.0mm}
\begin{figure}[t]
\begin{picture}(80,75)(0,1)
\mbox{\psfig{figure=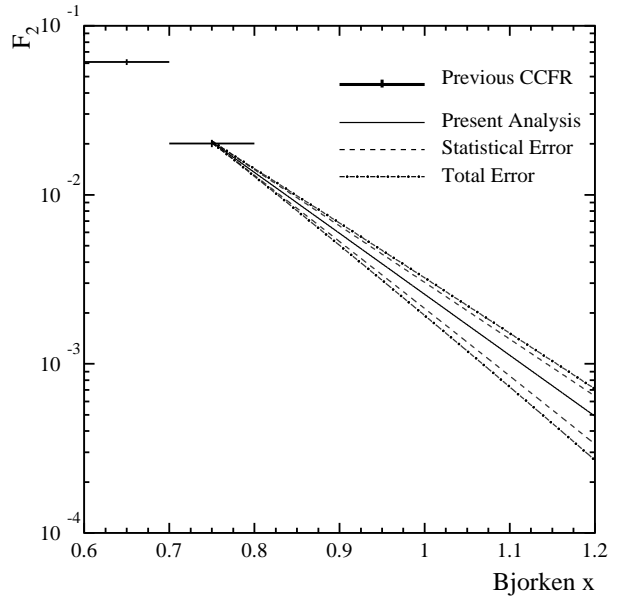,width=8.5cm,bbllx=.25in,bblly=2.25in,bburx=7.5in,bbury=9.25in}}
\end{picture}
\caption{
The exponentially falling $F_2$ structure function for $Q^2=125
(GeV/c)^2$ along with
the last two points from the previous CCFR analysis.  The solid 
line corresponds 
to our best fit to an exponetial $F_2$ structure function.  The area included 
by the dashed lines  corresponds to the area $\pm 1 \sigma$ statistical
error around 
the best fit.  The dotted-dashed lines indicate
the region allowed by adding both 
statistical and systematic errors in quadrature.  The normalization
of this graph is taken from all the data with $0.6 \le x \le 0.7$.  The
CCFR points plotted are only for the data $100\ ({\rm GeV/c})^2 \le Q^2
\le 150\ ({\rm GeV/c})^2$.
}
\label{fig:f2}
\end{figure}

We have investigated various systematic effects that might affect our
determination of the exponential slope parameter.  These are given in 
Table 1.  The biggest effect comes from our assumed muon energy resolution. 
If we broaden or narrow the resolution function from the measured 
one by one standard deviation, we can change the value of the 
slope parameter by $\pm 0.6$.
Other systematic effects, given in Table 1, lead to a total systematic
error on the slope parameter of $\pm 0.7$.  Thus our measured exponential
slope parameter is 
\begin{equation}
s=8.3\pm 0.7({\rm stat.})\pm 0.7 ({\rm sys.}).
\end{equation}

\begin{table}[t]
\begin{tabular}{c c} 
{\bf Systematic Factor} & {\bf Error in $s$}  \\ \hline
 Energy scale  &  0.2 \\ 
 Relative calibration & 0.1\\ 
 Incoming neutrino angle & 0.1 \\ 
 Outgoing muon angle    & 0.1 \\ 
 Hadronic energy mismeasurement & 0.1 \\ 
 Radiative corrections  & 0.1 \\ 
 Muon resolution function  & 0.6 \\ \hline
TOTAL SYSTEMATIC ERROR  & 0.7 \\ 
\end{tabular}
\caption{Systematic errors.}
\label{tab:system}
\end{table}

Other experiments have measured the expontential slope parameter $s$
for nuclear targets in the high $x$ region. 
Experiment E133\cite{SLAC} at SLAC measured the low energy 
electron-aluminum DIS cross section. 
Their value of $s= 7-8$ cannot be directly compared with 
our result since their high $x$ events come from the resonance region at 
low $Q^2$ and
not from the DIS region.
The muon-carbon data from BCDMS comes from the DIS region and should be 
comparable to ours.
They find a value of  $s=16.5 \pm 0.5$ (much steeper slope).  
This may indicate that the inter-nucleon effects
are much greater for our iron target than for their carbon target, although
this is not expected theoretically\cite{private}. 
Recent studies at lower energies\cite{somebody} from SLAC-NE3 and CEBAF E89-008
have suggested a consistent nuclear
picture if $\xi = {2 x /( 1 + \sqrt{1+{ 4 m_p^2 x^2\over Q^2}})}$ 
scaling is used.  However, their exponent of approximately
$s=17$ at $x=1$ is inconsistent with our data 
(in our $Q^2$ regime, $\xi \approx x$).  
Our result agrees with the theoretical prediction of $s=8-9$ by Strikman
and Frankfurt and is larger than the Baldin's prediction\cite{baldin}
of $s=6$.

The model that we used in this analysis assumed a
$Q^2$ dependence of the structure function of $F_2(x,Q^2) \propto 
(Q^2)^{-0.18}$.  This dependence is completely compatible with our data.

We
thank the staff of the Fermi National Accelerator Laboratory for
their hard work in support of this experimental effort.  We also wish
to thank Mark Strikman and Don Geesaman for their input on theoretical
issues.  This research was supported by the
U.S. Department of Energy and the National Science Foundation.


\end {document}